\documentclass[conference]{IEEEtran}
\IEEEoverridecommandlockouts
\usepackage[table]{xcolor} 

\usepackage{cite}
\usepackage{amsmath,amssymb,amsfonts}
\usepackage{algorithmic}
\usepackage{graphicx}
\usepackage{textcomp}
\usepackage{xcolor}
\usepackage{threeparttable}
\usepackage{tikz}

\def\BibTeX{{\rm B\kern-.05em{\sc i\kern-.025em b}\kern-.08em
    T\kern-.1667em\lower.7ex\hbox{E}\kern-.125emX}}

\usepackage{glossaries} 
\newacronym{FPGA}{FPGA}{Field Programmable Gate Array}
\newacronym{SVM}{SVM}{Support Vector Machines}
\newacronym{MF}{MF}{Match Filter}
\newacronym{FNN}{FNN}{Feed Forward Neural Network}
\newacronym{RFSoC}{RFSoC}{Radio Frequency System-On-Chip}
\newacronym{ADC}{ADC}{Analog-to-Digital Converter}
\newacronym{FTQC}{FTQC}{Fault-tolerant Quantum Computing}
\newacronym{I}{I}{in-phase}
\newacronym{Q}{Q}{quadrature}
\newacronym{PS}{PS}{Processing System}
\newacronym{PL}{PL}{Programmable Logic}
\newacronym{MAC}{MAC}{Multiply-Accumulate}
\newacronym{CNN}{CNN}{Convolutional Neural Networks}
\newacronym{RNN}{RNN}{Recurrent Neural Networks}
\makeglossaries

\DeclareRobustCommand*{\IEEEauthorrefmark}[1]{%
  \raisebox{0pt}[0pt][0pt]{\textsuperscript{\footnotesize\ensuremath{#1}}}}

\begin{document}

\newcommand\copyrighttext{%
  \footnotesize \textcopyright 2025 IEEE. Personal use of this material is permitted.
  Permission from IEEE must be obtained for all other uses, in any current or future
  media, including reprinting/republishing this material for advertising or promotional
  purposes, creating new collective works, for resale or redistribution to servers or
  lists, or reuse of any copyrighted component of this work in other works.}
  
\newcommand\copyrightnotice{%
\begin{tikzpicture}[remember picture,overlay]
\node[anchor=south,yshift=10pt] at (current page.south) {\fbox{\parbox{\dimexpr\textwidth-\fboxsep-\fboxrule\relax}{\copyrighttext}}};
\end{tikzpicture}%
}

\title{KLiNQ: Knowledge Distillation-Assisted Lightweight Neural Network for Qubit Readout on FPGA
}

\author{\IEEEauthorblockN{Xiaorang Guo\IEEEauthorrefmark{*,1}, Tigran Bunarjyan\IEEEauthorrefmark{*,1}, Dai Liu\IEEEauthorrefmark{1}, Benjamin Lienhard\IEEEauthorrefmark{2,3} and Martin Schulz\IEEEauthorrefmark{1}}

\IEEEauthorblockA{\IEEEauthorrefmark{1}Chair of Computer Architecture and Parallel Systems, Technical University of Munich, Garching, Germany \\ 
\IEEEauthorrefmark{2}Department of Chemistry, Princeton University, Princeton, NJ 08544, USA\\
\IEEEauthorrefmark{3}Department of Electrical and Computer Engineering, Princeton University, Princeton, NJ 08544, USA\\
Email: \{xiaorang.guo, tigran.bunarjyan, dai.liu\}@tum.de, blienhard@princeton.edu, schulzm@in.tum.de}

\thanks{\IEEEauthorrefmark{*} Equal Contribution.

This work was funded by the German Federal Ministry of Education and
Research (BMBF) under the funding program Quantum Technologies - From
Basic Research to Market under contract number 13N16087, as well as from
the Munich Quantum Valley (MQV), which is supported by the Bavarian State
Government with funds from the Hightech Agenda Bayern.}
}

\maketitle
\copyrightnotice
\begin{abstract}
Superconducting qubits are among the most promising candidates for building quantum information processors. Yet, they are often limited by slow and error-prone qubit readout---a critical factor in achieving high-fidelity operations. While current methods, including deep neural networks, enhance readout accuracy, they typically lack support for mid-circuit measurements essential for quantum error correction, and they usually rely on large, resource-intensive network models. This paper presents KLiNQ, a novel qubit readout architecture leveraging lightweight neural networks optimized via knowledge distillation. Our approach achieves around a 99\% reduction in model size compared to the baseline while maintaining a qubit-state discrimination accuracy of 91\%. KLiNQ facilitates rapid, independent qubit-state readouts that enable mid-circuit measurements by assigning a dedicated, compact neural network for each qubit. Implemented on the Xilinx UltraScale+ FPGA, our design can perform the discrimination within 32~$ns$. The results demonstrate that compressed neural networks can maintain high-fidelity independent readout while enabling efficient hardware implementation, advancing practical quantum computing.
\end{abstract}

\glsresetall
\section{Introduction}
Quantum computers are expected to tackle specific complex problems much faster than classical computers by leveraging quantum-mechanical properties. Among various quantum bit (qubit) technologies, superconducting qubits stand out for their fast gate operations and high fidelity. However, within the control pipeline of superconducting quantum computers, qubit readout remains one of the slowest and most error-prone processes~\cite{Zheng24,Walter17}. In this case, improving the qubit readout accuracy helps to enable high-fidelity quantum computers. Accelerating the readout process is essential for supporting the operation of large quantum circuits within the limited coherence times. 

Qubit readout is a complex process that involves both analog and digital operations. In this work, we focus on the digital aspect, processing the \gls{I} and \gls{Q} signals obtained from \glspl{ADC} to discriminate between the qubit states ('0' and '1'). This process is also known as single-shot qubit-state discrimination~\cite{Benjamin2022,stefanazzi2022qick}. In order to achieve high-fidelity qubit readout, a robust and accurate discrimination algorithm is required. Previous works have applied simple machine learning or statistical techniques, such as \glspl{SVM}~\cite{SVM}, hidden Markov models~\cite{Markov}, and \glspl{MF}~\cite{MF}, to improve the readout fidelity. However, the methods still fall short of the requirements for high-fidelity quantum computing. Thus, Benjamin~et~al.~\cite{Benjamin2022} proposes a deep neural network-based approach that significantly enhances readout accuracy. This method utilizes the \gls{FNN} architecture and directly takes time-sequenced digitized signals from the \glspl{ADC} as input, without requiring post-processing. Once trained, it classifies the states of all qubits with high accuracy. 

Although the \gls{FNN} effectively discriminates the qubit states, the overhead of software-based classification poses a significant challenge for superconducting quantum computers. Since the readout signals must be transferred repeatedly between the control and readout hardware and software, the resulting latency often exceeds the qubit coherence time~\cite{Guo2023}. Therefore, \textit{readout discriminators must be implemented on dedicated hardware}.

For this reason, current designs are migrating neural networks onto \glspl{FPGA} to reduce both communication and discrimination latency. The state-of-the-art work named HERQULES~\cite{Maurya2023} introduces an architecture that integrates specific \glspl{MF} tailored to each qubit into the \gls{FNN} proposed by work~\cite{Benjamin2022} but with a reduced network size benefiting from their selected features. This architecture improves the readout accuracy; however, several critical challenges remain unresolved: 1) the current \gls{FNN} architecture requires simultaneous readout of all qubits, which limits its applicability in mid-circuit measurement scenarios where qubits need to be measured potentially at arbitrary times and their outcomes influence subsequent operations; 2) the design requires an additional digital demodulation process, hindering real-time processing performance; and 3) existing designs are either too heavy to scale to large systems or are feature-limited for independent qubit processing. For instance, when HERQULES is adapted for single-qubit readout, its performance significantly degrades compared to its original configuration optimized for a five-qubit system. Another approach~\cite{gautam2024low} applies quantization techniques to reduce the \gls{FNN} size from work~\cite{Benjamin2022}. While this avoids demodulation, it sacrifices accuracy and fails to support mid-circuit measurements.


To address these challenges, we propose \textit{KliNQ: Knowledge Distillation-Assisted Lightweight Neural Network for Qubit Readout on FPGA}. KliNQ leverages knowledge distillation, where a large, feature-rich teacher network transfers its expertise to a compact student network optimized for FPGA implementation. By utilizing only the distilled knowledge, we reduce the network size by 99\% from the teacher model while maintaining a comparable discrimination accuracy of around 91\%. This significantly lowers \gls{FPGA} resource utilization, making KliNQ ideal for realizing \gls{FTQC} applications. In addition to shrinking the network, KliNQ employs an independent architecture with lightweight neural networks tailored to individual qubits. This design enables independent qubit discrimination without requiring simultaneous readouts, making KliNQ particularly well-suited for mid-circuit measurement and real-time feedback control.
Overall, our main contributions are:
\begin{itemize}
\item We introduce KliNQ, a qubit-state discriminator that leverages knowledge distillation to achieve around 99\% reduction in network size without compromising readout accuracy.
\item We design KliNQ with distilled independent neural networks for each qubit, enabling support for mid-circuit measurements while achieving a comparably high readout accuracy of 0.906 (average for five qubits).
\item We implement this design on an FPGA platform, utilizing a highly parallelized architecture to achieve readout times of 32~ns with low resource utilization.
\end{itemize}

\section{Backgrounds}
In this section, we discuss the principles of qubit readout for superconducting qubits and introduce knowledge distillation as a method for network compression.
\subsection{Qubit Readout}
In quantum computing, the qubit is the fundamental unit of storing information, capable of representing both states '0' and '1' in superposition, which is expressed by the formula \( |\psi\rangle = \alpha |0\rangle + \beta |1\rangle \), where \( |\alpha|^2 + |\beta|^2 = 1 \). During measurement, the qubit collapses to either '0' or '1', losing its superposition property. This process is known as qubit readout. For superconducting qubits, the readout is typically performed by sending microwave pulses into the qubit resonators for a few hundred nanoseconds to microseconds~\cite{gunyho2024single,Elisa2024}. Depending on the qubit state, the microwaves passing through the resonators undergo qubit-state-specific phase shifts, which are then down-sampled and eventually digitized by \glspl{ADC} into \gls{I} and \gls{Q} values. These \gls{I}/\gls{Q} values fluctuate over time during the readout, and their trace follows different convergence paths based on the qubit state, as introduced by Maurya~et~al.~\cite{Maurya2023}. Here, we focus on using neural networks to analyze these traces to discriminate the qubit state with high accuracy and as early as possible.

\subsection{Knowledge Distillation}
Knowledge distillation is a model compression technique that transfers knowledge from a large, complex neural network (the “teacher”) to a smaller, more efficient “student” network, enabling the creation of lightweight models that retain the predictive power of their larger counterparts while significantly reducing computational and memory demands~\cite{hinton2015distilling,cho2019efficacy}. The training process combines standard supervised learning with a distillation objective, aligning the student’s outputs with the teacher’s softened probabilities, which carry nuanced class relationship information~\cite{gou2021knowledge}. Since only lightweight student networks are needed during inference, this method proves highly effective in resource-constrained environments like FPGAs, where real-time performance and efficiency are critical. Additionally, knowledge distillation is advantageous for tasks requiring complex pattern recognition, such as qubit-state readout in quantum computing, facilitating models that maintain high accuracy while optimizing for reduced resource consumption and scalability.

\section{Methodology}
\label{sec:NN}
\begin{figure}[bt]
   \centering
\includegraphics[page=1,width=.4\textwidth]{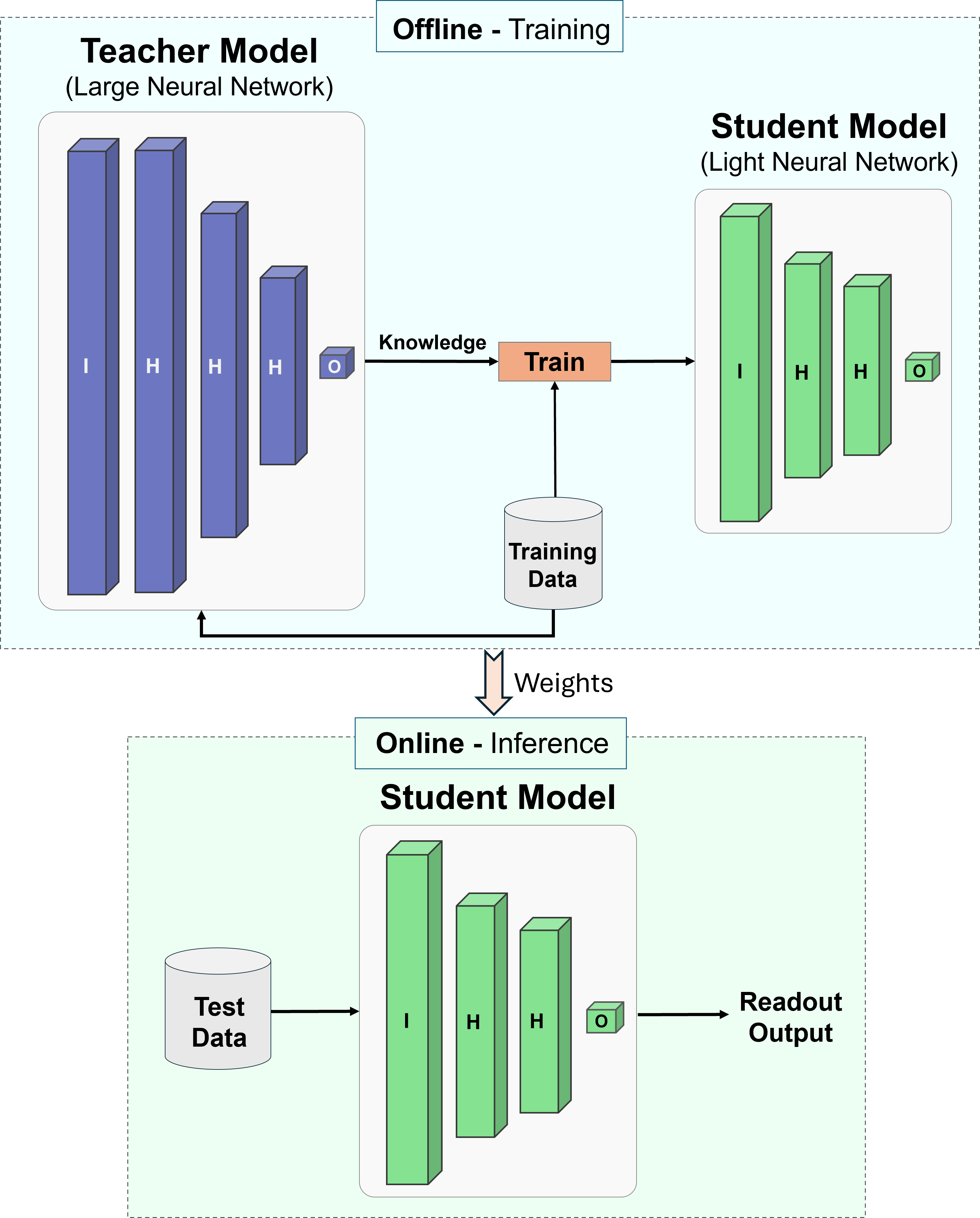}

 \caption{System Architecture of Knowledge Distillation-Assisted Discriminator: Offline Training and Efficient FPGA-Based Online Inference}
 \label{fig:sys_arc}
\end{figure}
Fig.~\ref{fig:sys_arc} illustrates the proposed system architecture of this work, which incorporates an offline training and distillation scheme alongside an online inference process. This section will discuss how knowledge distillation is generally performed in the system and the optimization strategies employed through hardware-software co-design to enhance performance and lead to the optimal model for KLiNQ. 
\subsection{Teacher Model: Experimental Design}
To develop a robust teacher model for knowledge distillation, we conduct extensive experiments to identify an architecture that effectively balances accuracy, generalizability, and efficiency. Prior research~\cite{Benjamin2022, Maurya2023, gautam2024low} has shown that \gls{FNN} outperforms \gls{CNN} and \gls{RNN} for qubit-state readout tasks. Compared to more complex architectures, the simplicity and flexibility of \gls{FNN} enable efficient processing of I/Q traces and ensure robust generalization, making it the ideal choice for this application.

The teacher \gls{FNN} was trained on 1~$\mu$s qubit measurement data using multiplexed I/Q signal traces flattened into 1000 inputs. The network architecture includes three hidden layers with 1000, 500, and 250 neurons, followed by an output layer for binary classification to distinguish between qubit states ('0' or '1'), optimizing fidelity scores across all qubits. We used the geometric mean of individual qubit assignment fidelities to evaluate the network models as the primary metric for discrimination accuracy. This geometric mean fidelity, $F_{\text{GM}}$, is defined as:

\[
F_{\text{GM}} = \left( \prod_{i=1}^N F_i \right)^{1/N},
\]

where $N$ represents the total number of qubits and $F_i$ is the readout accuracy for each qubit. This metric emphasizes consistent performance by penalizing outliers with low accuracy, ensuring reliable model performance across all qubits.

\subsection{Data Preprocessing and Input Optimization}
To ensure that the student networks achieve high fidelity and computational efficiency from a hardware-software codesign perspective, we need to optimize the input representation as small as possible. This involves two key steps: (1) averaging measurements over carefully selected intervals and (2) incorporating matched filters to enhance signal quality.

\subsubsection{Averaging Interval Values}
We perform a series of steps to preprocess the qubit-specific traces to achieve a compact input size while maintaining classification accuracy. We experimentally determine the optimal measurement interval for averaging I and Q components for each trace, effectively reducing the trace duration and dimensionality. Averaging compresses the raw data into a lower-dimensional representation by summarizing temporal features while preserving the most salient characteristics essential for classification. 

\subsubsection{Matched Filters}
While averaging significantly reduced the input size, experiments show that averaged trace alone cannot achieve a high classification fidelity, especially for qubits with subtle qubit-state-readout signal differences. To address this, we incorporate \glspl{MF} to provide an additional feature that enhances the discrimination capability of the student models. The MF envelope for each qubit was trained by maximizing the separation between ground-state ('0') and excited-state ('1') traces. The envelope is computed as:

\[
\text{MF Envelope} = \frac{\text{mean}(T_0 - T_1)}{\text{var}(T_0 - T_1)},
\]

where \(T_0\) and \(T_1\) represent the readout traces for the ground and excited states, respectively. During inference, the MF output is calculated by applying the trained envelope to the respective trace via a dot product operation, resulting in a single scalar feature. This scalar value, along with averaged I and Q components, is used as an additional input feature to form a representative input for the student models. This has been proven to significantly improve the performance of the student models by enhancing the signal-to-noise ratio (SNR) and capturing qubit-specific dynamics.

\begin{figure}[tb]
   \centering
\includegraphics[page=1,width=.48\textwidth]{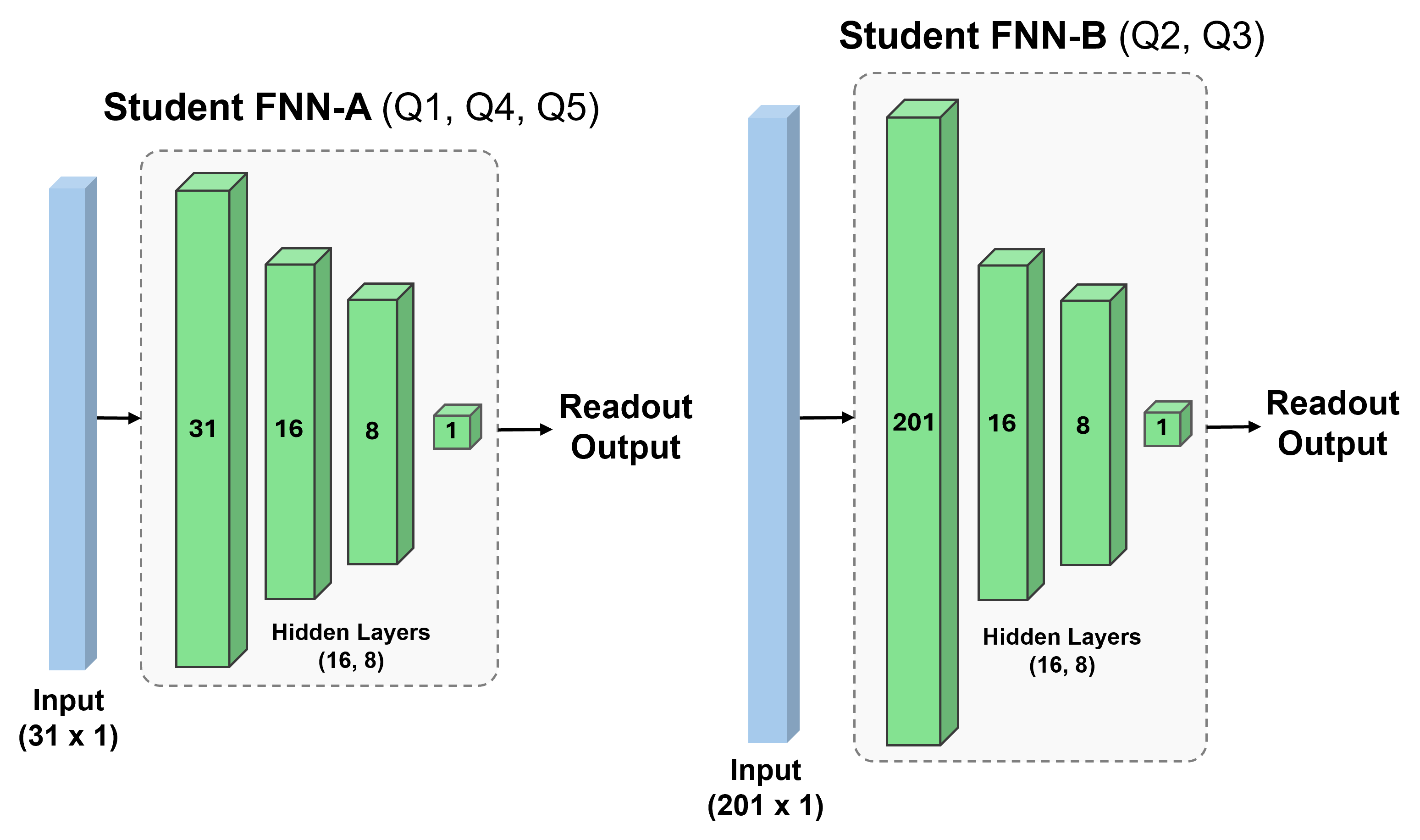}
 \caption{KLiNQ Student Network architectures for qubit groups of (Q1, Q4, Q5) and (Q2, Q3).}
 \label{fig:stu-fnn}
\end{figure}

\subsection{Distillation Training and Process}
Knowledge distillation is implemented to transfer the predictive capabilities of the teacher network to smaller, more efficient student networks. The primary objective of this process is to ensure that the student models maintain high classification accuracy while achieving a compact architecture, making them suitable for resource-constrained environments.

\textbf{Distillation Training:} The training process involves teaching the student network to replicate the behavior of the teacher network by minimizing a composite loss function. This loss combines two components: 
(1) a supervised learning objective, ensuring the student learns from the dataset’s ground truth labels (hard labels) and 
(2) a distillation objective, aligning the student’s outputs with the teacher’s predictions (soft labels). The composite loss function, \(\mathcal{L}_{\text{distill}}\), is defined as:

\[
\mathcal{L}_{\text{distill}} = \alpha \mathcal{L}_{\text{CE}} + (1 - \alpha) \mathcal{L}_{\text{KD}},
\]

where \(\mathcal{L}_{\text{CE}}\) represents the binary cross-entropy loss, calculated between the student’s predictions and the ground truth (hard labels).\(\mathcal{L}_{\text{KD}}\) stands for the distillation loss, computed as the Mean Squared Error (MSE) between the softened logits of the teacher and the student (soft labels), and \(\alpha\) represents a weighting factor that balances the supervised learning and distillation components.


\begin{figure*}[htbp]
   \centering
\includegraphics[page=1,width=.75\textwidth]{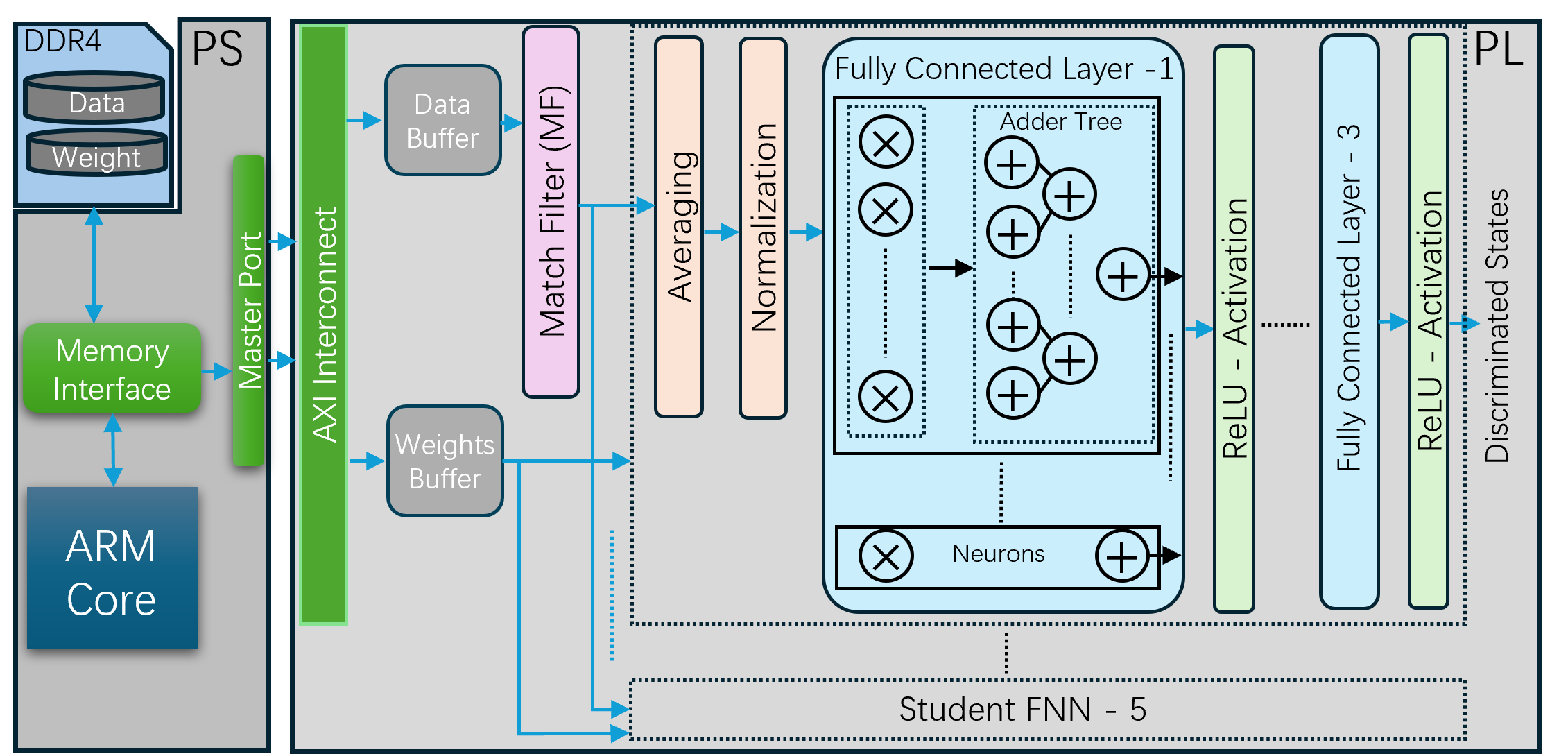}
 \caption{FPGA-based FNN architecture.}
 \label{fig:fpga-pipe}
\end{figure*}

\textbf{Distillation Process and Inference:} As mentioned earlier in Fig.~\ref{fig:sys_arc}, distillation happens during the training phase, where the teacher model is pre-trained on the dataset and subsequently used to guide the student model. The pre-trained teacher network provides softened probabilities (soft labels) as supplementary supervisory signals for the student network's distillation process. The student network, initialized with random weights, is trained using gradient descent to minimize a composite \(\mathcal{L}_{\text{distill}}\).

For inference, only the student models are deployed, having inherited the predictive capabilities of the teacher network through offline distillation. The compact design of the student models supports efficient, low-overhead inference, making them ideal for qubit-state discrimination in subsequent \gls{FPGA} implementations.

\subsection{Student Network Architectures}
\label{sec:NN_student}
We now introduce two distinct student network architectures, as illustrated in Fig.~\ref{fig:stu-fnn}, based on the experimental observations: a small \gls{FNN}-A for qubits 1, 4, and 5, which exhibit high SNR and are easier to discriminate, and a relatively larger \gls{FNN}-B for qubits 2 and 3, where the SNR is suboptimal, particularly for qubit 2. Note that the input size of the networks is fixed, and when the trace length changes, we dynamically adjust the number of samples to be averaged to match the required output size. In this section, we use 1~$\mu s$-long traces as an example for description.


\begin{itemize}
    \item \textbf{Student FNN-A (Q1, Q4, Q5):}
    This architecture is optimized for qubits whose qubit-state discrimination performance is adequately captured by the 64-$ns$ (32 samples) averaging interval. The input consists of the 30-dimensional averaged I/Q data plus the \gls{MF} feature, forming a 31-dimensional vector. The network comprises two hidden layers with 16 and 8 neurons, respectively, followed by a single output layer for binary classification.
    
    \item \textbf{Student FNN-B (Q2, Q3):}
    The state discrimination of qubits 2 and 3 requires larger input sizes due to increased noise levels, crosstalk effects, and greater variability in the readout traces. For these qubits, a 10-$ns$ (5 samples) averaging interval has been identified as optimal, resulting in 200 averaged data points from the I/Q traces. The input size was expanded to 201 dimensions when combined with the MF feature. This network also features two hidden layers with 16 and 8 neurons, followed by a single output layer.
\end{itemize}

These architectures have been selected as the most optimal configurations in 
terms of balancing fidelity, computational efficiency, and input sizes.

\section{Lightweight Neural Networks on FPGA}

In this section, we describe the deployment of our lightweight student neural network on FPGA and discuss its architecture. As mentioned in Sec.~\ref{sec:NN}, due to the different properties of qubits, we design different sizes of neural networks to balance accuracy and efficiency. In this case, the network we develop on FPGA is also re-configurable in size. 

Fig.~\ref{fig:fpga-pipe} illustrates the overall structure of the FPGA-based FNN, which comprises DDR memory, the \gls{PS} part, and the \gls{PL} component. The DDR memory is responsible for storing the test data of the qubit traces as well as the trained weights of the neural network. To be noted, in actual quantum systems, KliNQ can process the raw data directly from the \glspl{ADC} to perform state discrimination. However, this work focuses on the algorithm and FPGA implementation, so we use the DDR memory to store the traces as a substitute. \gls{PS} serves as a data exchange interface between the memory and our customized logic to offload the qubit traces, necessary parameters, and weights to \gls{PL} through the AXI bus protocol.

In the \gls{PL}, we adopt a modular design approach to deploy each \gls{FNN} layer sequentially, enabling pipelined functionality. As discussed in Sec.~\ref{sec:NN_student}, two network architectures are developed for different qubits to optimize performance. Although these models differ in configurations, they share the same overall architecture. Therefore, rather than introducing them separately, we provide a unified overview. We select a 32-bit fixed-point format for the data representation, allocating 16 bits for the integer and 16 bits for the fractional part to maintain discrimination accuracy.

Following the architectures outlined in Sec.~\ref{sec:NN_student}, the data pre-processing stage begins with an average layer that computes the average value of a group of qubits, reducing the input size to the neural network and minimizing resource utilization. This is followed by a normalization layer, which optimizes the data distribution, improves algorithm efficiency, and mitigates the risk of overflow in the fully connected layers, where multiple multiplication and addition operations are performed. To perform normalization, the expression \( \frac{x - x_{\text{min}}}{\sigma_x} \) is used in this work, where $\sigma_x$ represents the standard deviation of the observed qubit traces. The original formula typically requires division, which consumes significant resources and time on the FPGA, limiting real-time performance. However, as the $x_{\text{min}}$ and $\sigma_x$ are prepared during training, the $\sigma_x$ is also approximated as a power of 2. Thus, we replace the division with shift operations and can get the results within only two clock cycles. 
Simultaneously, another feature is extracted from the \gls{MF} layer. Given the identical operations and data flow (\gls{MAC}), this layer reuses the same design as a fully connected layer, which will be briefly introduced later.

After normalization, we concatenate the \gls{I} and \gls{Q} values together with the \gls{MF} feature into a single input vector, which is then fed into the fully connected layer. Within each layer, neuron operations are executed in parallel, so the latency of a layer equals that of a single neuron. Neurons perform input-weight multiplications in a 4-stage pipeline, utilizing time-multiplexed DSP resources on the FPGA, which are exclusively responsible for executing these multiplication operations. Each multiplication is implemented using fully combinational logic, achieving a latency of one clock cycle, so the overall latency of multiplication is four cycles. The outputs are summed along with the bias through an adder tree structure (shown in Fig.~\ref{fig:fpga-pipe}), achieving a latency with  $\lceil \log_2(\text{n}) \rceil + 1$, where \( n \) is the number of inputs. However, the latency varies slightly with the number of inputs due to the ceiling operations.
Each fully connected layer is followed by a ReLU activation function, which introduces the non-linearity between layers. The sign bit is checked to determine whether the output remains unchanged or is set to zero. Additionally, overflow conditions are managed in the activation layer to ensure correct functionality. Meanwhile, we handle overflows in the activation layer to ensure the correct functionality. Subsequently, the whole process is repeated twice more, culminating in a final single value that serves as the discrimination result.

\section{Evaluations}
\subsection{Experimental Setup}
\textbf{Datasets and baseline:} In this work, we train and evaluate our method using real-measured experimental data from a superconducting quantum information processor comprising five qubits~\cite{Benjamin2022}. The dataset comprises measurements from all 32 possible qubit-state permutations, where each qubit's state is characterized by a sequentially recorded \gls{Q} and \gls{I} value. While the full recorded trace spans 2 $\mu s$, only the first 1 $\mu s$ is utilized in our analysis, as this duration provides nearly identical discrimination accuracy compared to the full sequence. For network evaluation, we adopt a setup similar to previous works~\cite{Benjamin2022, Maurya2023}, leveraging 15,000 traces per qubit-state configuration for training and 35,000 traces for testing. In addition, we use work~\cite{Benjamin2022} as the baseline model in the comparison.

\textbf{FPGA Hardware:} The hardware implementation of this design is programmed in Verilog and synthesized using Xilinx Vivado 2024. The complete project is deployed on a Xilinx Zynq RFSoC (ZCU216)~\cite{rfsoc}, operating at 100~MHz. 

\subsection{Discrimination Accuracy}
We begin by evaluating one key metric for discriminators: readout fidelity. Tab.~\ref{table:klinq_comparison} presents the readout fidelities of KLiNQ compared with related works with the readout-trace length of 1 $\mu s$. The comparison includes the individual readout accuracies and their geometric means of the readout fidelities for five qubits. To provide a comprehensive comparison, we compute two geometric means: one for all five qubits and another excluding qubit 2, as noise significantly impacts its performance. The original papers of the baseline \gls{FNN} and HERQULES are designed for a five-qubit synchronous readout system. However, in this comparison, we reproduce their design and test their architecture on independent readouts. The result shows that KLiNQ achieves a competitive fidelity compared to the large baseline model and outperforms HERQULES with more than 1\%. Here, we exclude work~\cite{gautam2024low} as it does not explicitly report accuracy. 

\begin{table}[htbp]
\caption{Qubit-Readout Fidelity Comparison Tested on Independent Readout Scenario}
\label{table:klinq_comparison}
\centering
\resizebox{\linewidth}{!}{%
\begin{threeparttable}
\begin{tabular}{|l|c|c|c|c|c|c|c|}
\hline
\textbf{Design} & \textbf{Qubit 1} & \textbf{Qubit 2} & \textbf{Qubit 3} & \textbf{Qubit 4} & \textbf{Qubit 5} & \textbf{F$_{5Q}$} & \textbf{F$_{4Q}$} \\ \hline
Baseline FNN\tnote{1} & 0.969 & 0.748 & 0.940 & 0.946 & 0.970 & 0.910 & 0.956 \\ \hline
HERQULES\tnote{2} & 0.965 & 0.730 & 0.908 & 0.934 & 0.953 & 0.893 & 0.940 \\ \hline
\textbf{KLiNQ} & 0.968 & 0.748 & 0.929 & 0.934 & 0.959 & 0.904 & 0.947 \\ \hline
\end{tabular}
\begin{tablenotes}
\item[1] Baseline FNN~\cite{Benjamin2022} reports a geometric mean (F$_{5Q}$) of 0.912 for the five-qubit readout system.
\item[2] HERQULES~\cite{Maurya2023} reports a geometric mean (F$_{5Q}$) of 0.927 for the five-qubit readout system.
\end{tablenotes}
\end{threeparttable}%
}
\end{table}

\begin{figure}[htbp]
   \centering
\includegraphics[width=.48\textwidth]{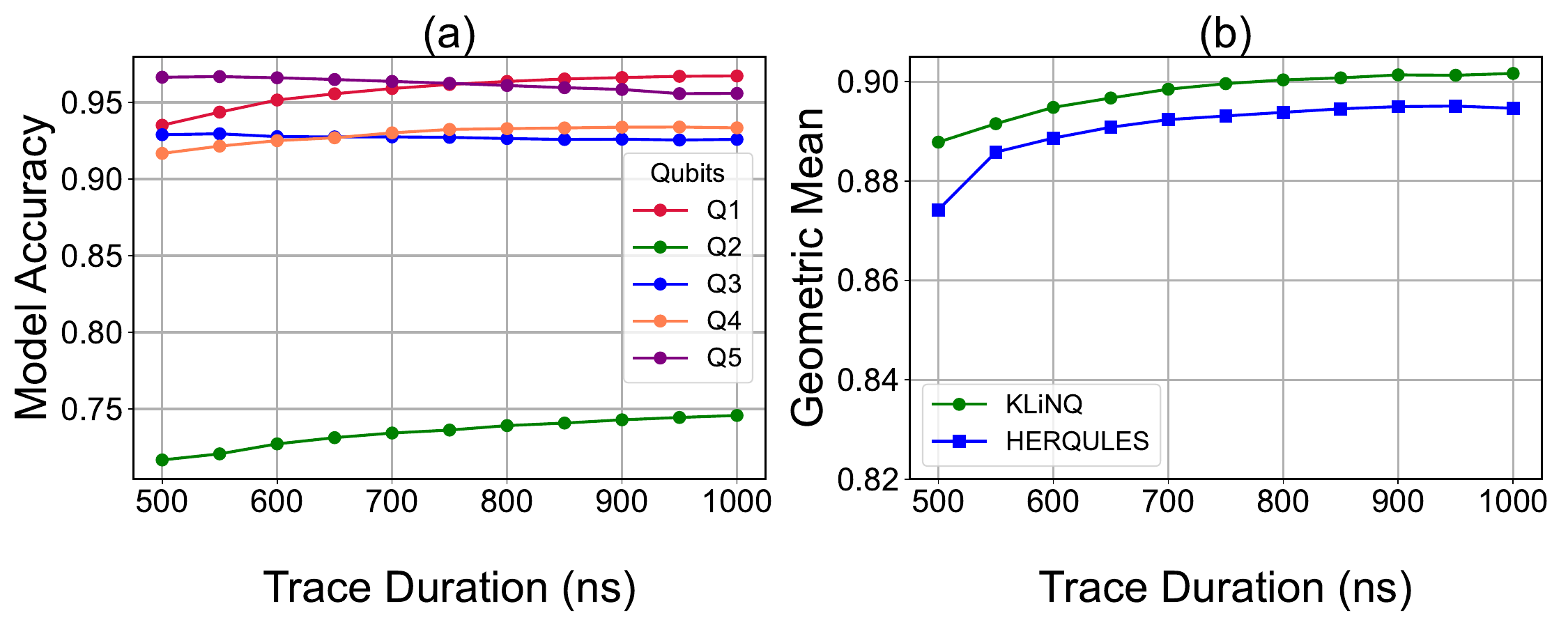}
 \caption{(a) Accuracy of individual qubit-state discriminator when varying the readout-trace duration. (b) Geometric mean comparison of readout fidelity between KLiNQ and HERQULES across varying readout-trace lengths.}
 \label{fig:gm_acc}
\end{figure}

We further evaluate the robustness of KLiNQ's readout performance across varying trace durations. Tab.~\ref{table:klinq_accross_trace_duration_acc} summarizes the readout fidelities as trace durations decrease from 1~$\mu$s to 500~ns. While the geometric mean fidelity across all five qubits declines with shorter traces, certain qubits achieve optimal performance at specific shorter durations, highlighted in green. By utilizing the optimal trace durations for each qubit, KLiNQ achieves an improved five-qubit fidelity (\(F_{5Q}\)) of 0.906. The behavior of the fidelities across trace lengths is also depicted in Fig.~\ref{fig:gm_acc}(a), where all qubits except Qubit 2 demonstrate consistent and robust behavior. Moreover, Fig.~\ref{fig:gm_acc}(b) shows that KLiNQ maintains superior readout fidelity and robustness compared to the state-of-the-art HERQULES, particularly excelling at shorter readout-trace durations.

\begin{table}[htbp]
\caption{Readout Fidelity of KLiNQ VS. Readout-Trace Duration}
\label{table:klinq_accross_trace_duration_acc}
\centering
\resizebox{\linewidth}{!}{
\begin{tabular}{|l|c|c|c|c|c|c|c|}
\hline
\textbf{Design} & \textbf{Duration} & \textbf{Qubit 1} & \textbf{Qubit 2} & \textbf{Qubit 3} & \textbf{Qubit 4} & \textbf{Qubit 5} & \textbf{F$_{5Q}$}  \\ \hline
 & 1$\mu$s  & \cellcolor{green!20}0.968 & \cellcolor{green!20}0.748 & 0.929 & 0.934 & 0.959 & 0.904 \\ \cline{2-8}
 & 950ns    & 0.967 & 0.744 & 0.925 & \cellcolor{green!20}0.934 & 0.956 & 0.901 \\ \cline{2-8}
\textbf{KLiNQ NNs} & 750ns    & 0.962 & 0.736 & 0.927 & 0.932 & 0.963 & 0.900  \\ \cline{2-8}
 & 550ns    & 0.944 & 0.720 & \cellcolor{green!20}0.930 & 0.921 & \cellcolor{green!20}0.967 & 0.891  \\ \cline{2-8}
 & 500ns    & 0.935 & 0.717 & 0.929 & 0.917 & 0.966 & 0.887  \\ \hline
\end{tabular}%
}
\end{table}




\label{sec:NN_accuracy}


\subsection{Network Compression Rate}

\begin{figure}[htbp]
   \centering
\includegraphics[width=.4\textwidth]{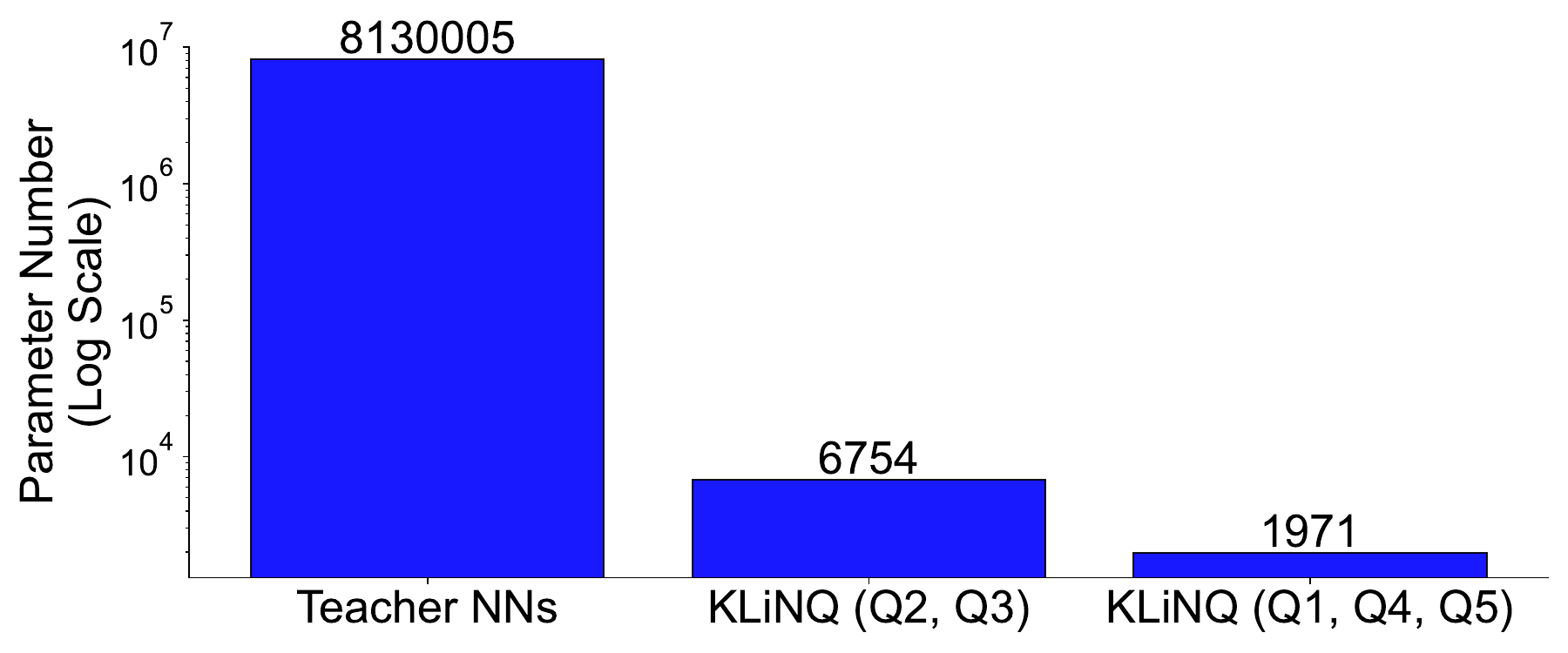}
 \caption{Comparison of the number of network parameters for the teacher model and two distilled KLiNQ Student NNs (the plot is in log scale).}
 \label{fig:compression_rate_histogram}
\end{figure}

Network Compression Rate (NCR) is a key metric in evaluating knowledge distillation. To assess this, we compare the number of parameters in KLiNQ's five individual qubit discrimination system to those in the teacher NN architectures. As illustrated in Fig.~\ref{fig:compression_rate_histogram}, the teacher NNs collectively comprise a substantial 8.14 million parameters. In contrast, KLiNQ's student NNs are remarkably compact, with the larger FNN (used for qubits 2 and 3) containing just 6,754 parameters, and the smaller FNN (used for qubits 1, 4, and 5) comprising only 1,971 parameters. This corresponds to an NCR of 99.89\% relative to the teacher neural network. Furthermore, compared to the baseline model with 1.63 million parameters, KLiNQ also achieves a remarkable 98.93\% reduction in network size, highlighting the exceptional efficiency of its knowledge distillation strategy.


\subsection{Latency and Resource Utilization}
Tab.~\ref{tab:resource} summarizes the FPGA resource utilization and model latency for the subcomponents of the proposed design, including the \gls{MF} for both I/Q components, the average and normalization module (AVG\&NORM) and the neural network. The \gls{MF} is time-multiplexed across all qubits, so its resource usage is counted only once for the entire design. In contrast, the AVG\&NORM module and neural networks are instantiated separately for each qubit, with either the small or large configuration selected based on the setup requirements.

The overall latency is calculated by summing the latencies of the pipelined components. Since the logarithmic values of the samples for both longer traces (1~$\mu s$) and shorter acceptable traces (550~$ns$) are the same, the latency remains constant across all readout traces. Additionally, while the larger model requires more time for inference, it benefits from shorter averaging groups, leading to a reduced average processing time. As a result, both modules coincidentally produce the same execution latency of 32~$ns$ at 100~MHz.

Resource utilization can be further optimized, as the current design prioritizes minimizing discrimination latency, sacrificing time-multiplexing to achieve faster classification. Therefore, trade-offs can be made to balance resource efficiency and performance depending on the specific latency requirements.
\begin{table}[h!]
\caption{Resource utilization and latency for individual components}
\label{tab:resource}
\centering
\resizebox{0.9\linewidth}{!}{%
\begin{tabular}{|c|c|c|c|c|}
\hline
\textbf{Components} & \textbf{LUT} & \textbf{FF} & \textbf{DSP} & \textbf{Latency ($ns$)} \\
\hline
\multicolumn{5}{|c|}{\textbf{Shared Resources}} \\
\hline
MF         & 27180 (6.39\%) & 24052 (2.83\%) & 375 (8.78\%)  & 11 \\
\hline
\hline
\multicolumn{5}{|c|}{\textbf{Per-Qubit Resources: Qubits 1, 4, 5}} \\
\hline
AVG\&NORM           & 17770 (4.2\%)            & 11415 (1.35\%)            & 0 (0\%)       & 9 \\
\hline
Network                  & 8840 (2.08\%)  & 6020 (0.71\%)  & 55 (1.28\%)   & 12 \\
\hline
\hline
\multicolumn{5}{|c|}{\textbf{Per-Qubit Resources: Qubits 2, 3}} \\
\hline
AVG\&NORM           & 19600 (5\%)             & 17500 (2\%)              & 0 (0\%)       & 6 \\
\hline
Network                  & 25882 (8.44\%) & 23172 (2.72\%) & 226 (5.29\%)  & 15 \\
\hline
\end{tabular}}
\end{table}

\section{Discussion}
This work proposes an independent readout scheme to enable mid-circuit measurements and error correction without synchronizing all readouts into a unified large neural network. Although it brings this advantage, it introduces significant challenges due to inherent quantum system characteristics. As highlighted in previous work by Benjamin~et~al.~\cite{Benjamin2022}, the crosstalk in frequency-multiplexed readouts and correlations between qubits are a major source of error. Consequently, separating the readouts without accounting for inter-qubit influences inevitably leads to a reduction in fidelity. This trade-off is also evident in our results, where the independent readout approaches always underperform compared to the large network for the five-qubit system. Yet, this step represents a necessary direction for advancing systems toward \gls{FTQC}. 

On the other hand, this fidelity sacrifice underscores the complexity of mitigating crosstalk in independent readout schemes. Thus, future work will focus on integrating crosstalk information into the teacher model, thereby enabling the student model to learn and compensate for these effects. Additionally, exploring masking techniques within large networks could serve as another effective strategy to minimize these errors. By incorporating these considerations, we aim to achieve improved fidelity in individual readouts, ultimately advancing the practicality of error-corrected quantum computation.
\section{Conclusion}
This paper introduces KLiNQ, a method leveraging knowledge distillation to train lightweight neural networks optimized for FPGA implementation. We propose an individual qubit-state readout discriminator capable of enabling mid-circuit measurements while maintaining high readout fidelity. Experimental results demonstrate that KLiNQ achieves an average qubit-state-discrimination accuracy of around 0.91, and sustains an accuracy above 0.9 even with a reduced 750~$ns$ readout trace. The FPGA-based implementation features low hardware cost and an ultra-fast latency of 32~$ns$. Overall, our work highlights the potential of network compression techniques for efficient, hardware-based qubit-state readout.


\bibliographystyle{IEEEtran}
\bibliography{ref}
\end{document}